\documentclass[epjc3,twocolumn]{svjour3} 
\usepackage[T1]{fontenc}
\smartqed  
\RequirePackage{mathptmx}      
\RequirePackage{flushend}
\RequirePackage[numbers,sort&compress]{natbib}
\RequirePackage[colorlinks,citecolor=blue,urlcolor=blue,linkcolor=blue]{hyperref}
\usepackage{graphics}
\usepackage{graphicx}
\usepackage{amssymb}
\usepackage{amsmath}
\usepackage{slashed}
\newcommand{\be}{\begin{eqnarray}}
\newcommand{\ee}{\end{eqnarray}}
\newcommand{\beq}{\begin{equation}}
\newcommand{\eeq}{\end{equation}}

\newcommand{\bemul}{\begin{multline}}
\newcommand{\eemul}{\end{multline}}

\begin{document}
\title{Neutrino interactions with ultralight axion-like dark matter}
\author{Mat\'ias M. Reynoso\thanksref{e1,addr1}  \and Oscar A. Sampayo\thanksref{e2,addr1}\and Agust\'in M. Carulli\thanksref{e3,addr1}}

%
\thankstext{e1}{e-mail: mreynoso@mdp.edu.ar}
\thankstext{e2}{e-mail: sampayo@mdp.edu.ar}
\thankstext{e3}{e-mail: amcarulli@mdp.edu.ar}



\institute{IFIMAR (CONICET-UNMdP) and Departamento de F\'isica, Facultad de Ciencias Exactas y Naturales, Universidad Nacional de Mar del Plata, Funes 3350, (7600) Mar del Plata, Argentina.\label{addr1}}
\date{Received: date / Revised version: date}
%
\maketitle
\begin{abstract}
In this work, we study the propagation of high energy neutrinos produced in extragalactic sources including the effect of a possible interaction with ultralight axion-like  particles (UALP) with a mass $m_a\sim 10^{-22} {\rm eV}$ as the constituents of dark matter (DM) under the assumption that their coupling to neutrinos is dominant. We compute the cross section and describe the propagation of a diffuse neutrino flux using transport equations for each mass eigenstate. This allows us to obtain the neutrino fluxes of the different flavors to be observed at the Earth with neutrino telescopes under different assumptions for the flavor composition emitted at the sources and for a normal ordering (NO) or an inverted ordering (IO) of the neutrino masses. If the coupling of neutrinos with UALPs is the same for all flavors ($g_{\nu a}$), we find that interactions change the flavor composition of neutrinos arriving on Earth for $g_{\nu a}\gtrsim 0.5\,{\rm GeV}^{-1}$, causing the electron(muon) flavor to dominate in the NO(IO) case for neutrino energies above $\sim 10^5\,{\rm GeV}$. Although current data on the flavor ratios suggest that interactions with UALP DM do not take place within the range of coupling studied (particularly in the NO case) more data is needed to improve the precision of the experimentally measured flavor composition.    
\end{abstract}
%
%
%
\section{Introduction}
\label{intro}
The detection of high energy neutrinos of astrophysical origin is providing a new insight into the sources themselves and it can also help to probe different scenarios and properties of neutrino related physics \cite{halzen2018}. In particular, neutrino inteactions with dark matter (DM) has been considered and their effects have been investigated under different assumptions \cite{barranco2011,aeikens2015,reynoso2016,desalas2016,arguelles2017,rasmussen2017,brdar2017,huang2018,pandey2018,farzan2018, choi2019,koren2019,murase2019,penacchioni2020}. 

One interesting possibility is that dark matter can be composed of ultralight particles, as this could help to alleviate problems of cold dark matter models regarding the overproduction of both substructure
in the galactic haloes and satellite dwarf galaxies that are not observed \cite{hu2000}. 
Such extremely light particles can be pseudoscalars which arise naturally in scenarios beyond the Standard Model (SM) based on string theories \cite{witten2006} and are known as ultralight axion-like particles (UALPs). In particular, these models predict the  
existence UALPs spanning a wide range of masses, $\sim 10^{-33}-10^{-10} {\rm eV}$ \cite{arvanitaki2009,cicoli2012}. {While the lightest UALPs could be associated to dark energy \citep{catena2007,panda2011}, the appropriate mass for viable DM candidates is $m_a\sim 10^{-22}-10^{-21}{\rm eV}$, which has the advantage of preserving the large-scale behavior of the standard cold dark matter models, but at the same time it suppresses the substructure at shorter scales due to a large de Broglie wavelength $\sim 1\,{\rm kpc}$ \citep{hu2000,niemeyer2020}.  }   

In the present work, we study the effects of neutrino
interactions with such particles during the propagation
from extragalactic sources assuming that their evolution with the redshift follows the star formation rate (SFR) accoding to Ref. \cite{madau2014}. We follow the 
procedure discussed on a previous work with a different
type of interaction \cite{reynoso2016}. The interaction vertex analysed in this work was considered previously in Ref.\cite{huang2018} assuming that UALPs are neutrinophilic, i.e., their interactions with the charged leptons can be neglected (see also Ref. \citep{shoemaker2013}). In these conditions, the stringent bounds on the coupling to electrons based on stellar evolution considarations for globular clusters $g_{ae}\lesssim 4\times 10^{-9}\rm{GeV^{-1}}$ \citep{viaux2013} do not affect the coupling to neutrinos. The corresponding  astrophysical and cosmological constraints that can be applied in the present context are much less restrictive, as discussed in Ref. \citep{huang2018}. In their work, they also described the effects of modified neutrino oscillations in baseline experiments such as DUNE  \citep{dune2015} due to an effective potential included in the Hamiltonian.  Our approach here is complementary, since we consider a flavor diagonal coupling $g_{\alpha\beta}=g_{\nu a}\delta_{\alpha\beta}$ and this leads only to a global phase that implies no modification to oscillations due to interactions with UALP DM, and therefore no effects to be observed in baseline experiments. 

 Nevertheless, neutrino-UALP scatterings can still play a role and imprint effects on a diffuse neutrino flux. We explore this possibility in the present work by solving a transport of each mass (or propagation) eigenstate, for which the neutrino mass is well defined. The surviving fluxes of flavored neutrinos at the Earth are recovered by superposition, and this can be compared with observational data. We consider the cases where the hierarchy of neutrino masses are accommodated following a normal ordering (NO) or an inverted ordering (IO), and we explore three different scenarios of dominant neutrino sources with different initial flavor compositions. We find that interactions produce a change in the flavor composition of neutrinos arriving on Earth for $g_{\nu a}\gtrsim 0.5\,{\rm GeV}^{-1}$ in comparison to the outcome expected with no interactions. In particular, the electron(muon) flavor comes to dominate in the NO(IO) case for neutrino energies above $\sim 10^5\,{\rm GeV}$. Additionally, a directional dependence on the neutrino flavor ratios can be observed, given that galactic DM is more abundant in directions closer to the galactic center. 

By comparing the flavor ratios integrated on the neutrino energy with available flavor measurements by IceCube\cite{heseICRC2019,hese2020}, the realization of the interaction effects studied seems unlikely, since the best fit values of the flavor ratios are not close to the predicted in the case of interactions. However, more data with more observation time and larger detectors are necessary to achieve a much higher precission in the flavor composition measurements \cite{icecubegen2,bustamante2021} and this will help to constrain definitely scenarios as the discussed here. 

The rest of this work is organized as follows. In the next section, we compute the cross sections for the neutrino-UALP interaction, and in Section \ref{sec:3propagation}, we describe the neutrino propagation first in extragalactic space and then through the galactic DM halo. In Section \ref{sec4:results}, we present the results obtained for the neutrino fluxes and flavor ratios on Earth, and finally in Section \ref{sec5:discussion}, we conclude with a discussion.

\section{Neutrino interactions with ultralight axion-like particles}
\label{sec:2interactions}
In the present work, we consider the interactions between neutrinos and UALPs corresponding to the following Lagrangian term (e.g. \cite{huang2018,irastorza2018}):
\begin{equation}
\mathcal{L}_{\nu_\alpha a}= -ig_{\alpha\beta}\partial_\mu a  \bar{\nu}_{\alpha}\gamma^\mu \gamma_5 \nu_\beta, \label{lagr1}
\end{equation} 
where $\alpha,\beta=\{e,\mu,\tau\}$. 
 {This effective interaction involves only the active neutrinos, and it is implicitly assumed that the UALP coupling to the charged leptons is not relevant, which is a characteristic feature of neutrinophilic models (e.g. \citep{huang2018,shoemaker2013,blennow2019}). 
Without going into details, we note that this kind of interactions appear in models such as the applied in Ref. \citep{baek2019}, 
where a new scalar doublet and a new global symmetry U(1) are included in order to give small 
Dirac masses to the neutrinos.  In this type of models, known as
neutrinophilic two-Higgs doublet models (nu2HDM),  it is possible to obtain the type interactions sought, i.e., with a strong coupling to the neutrinos. Although they were originally proposed
for the QCD axion [34], a similar approach could be suitable in our context for UALPs. We
also focus on a minimal scenario where the UALP has flavor-conserving couplings at
tree level, as we discuss below.} 

The interaction can be decomposed into terms corresponding to each mass eigenstate $\nu_i$, given that $\nu_\alpha=\sum_{i=1}^3 U_{\alpha i}\nu_i$, where $U_{\alpha i}$ are the elements of the unitary mixing matrix for neutrinos \cite{pdg2020,esteban2020}.
Defining $q:= \bar{p}+p$ as the momentum carried by the UALP $a$, we have that $(\partial_\mu \, a) =i q_\mu \, a$,
and making use of Dirac equation, it follows that
 \be 
    (\partial_\mu a )\bar{\nu}_i\gamma^\mu \gamma_5\nu_j &=& - i\left(m_i +  m_j\right) a\bar{\nu}_i\gamma_5 \nu_j.
 \ee
 We then consider the case of a {flavor-diagonal coupling, where we have $g_{\alpha\beta}=g_{\nu a}\delta_{\alpha\beta}$ for all the flavors}. Then, Eq. (\ref{lagr1}) can be rewritten as
 \be
 \mathcal{L}_{\nu a}= -ig_{\nu a}\sum_{ij} \left(m_i +  m_j\right)\sum_\alpha U^*_{\alpha i}U_{\alpha j}a\bar{\nu}_i\gamma_5 \nu_j,
 \ee
and since $\sum_{\alpha}U^*_{\alpha i}U_{\alpha j}= \delta_{ij}$, we have the following interaction terms for each massive neutrino $\nu_i$: 
\be
\mathcal{L}_{\nu_i a}&=& -ig_{i}\, a\bar{\nu}_i\gamma_5 \nu_i, \label{lagr2}
\ee 
where $g_{i}= 2 m_i g_{\nu a}.$ 

{Since the coupling is flavor-diagonal and flavor-universal, this leads to an effective potential proportional to the identity. When this is  added to the Hamiltonian to study the effects of neutrino oscillations in a DM background (e.g., as in Refs. \citep{huang2018,farzan2018}), it only contributes with global phase which does not modify the relative energies of the different massive neutrinos. Hence, as mentioned above, no effects due to oscillations in DM medium are expected in the present context. However, the effect of neutrino scattering with UALPs can still be relevant and lead to an absorption of the neutrino flux at a given energy, as was also considered for other interaction vertices in previous works \citep{barranco2011,farzan2014,reynoso2016,arguelles2017}. }

 {The fact that consider only flavor-diagonal and flavor-universal interactions also allows to avoid very strong constraints for neutrinos decays \citep{hannestad2005}. Still, these asumptions correspond to a possibility which is usually explored in similar studies (e.g. \citep{hannestad2005,huang2017}), and in particular, the only relevant effect is due to neutrino-DM scattering. Hence, if off-diagonal couplings are considered, this would bring into play the effects of neutrino oscillations in a DM medium, as discussed above, but the study of this possibility is left for future work. }

\begin{figure}
\includegraphics[width=0.75\textwidth,trim=120 590 0 150,clip]{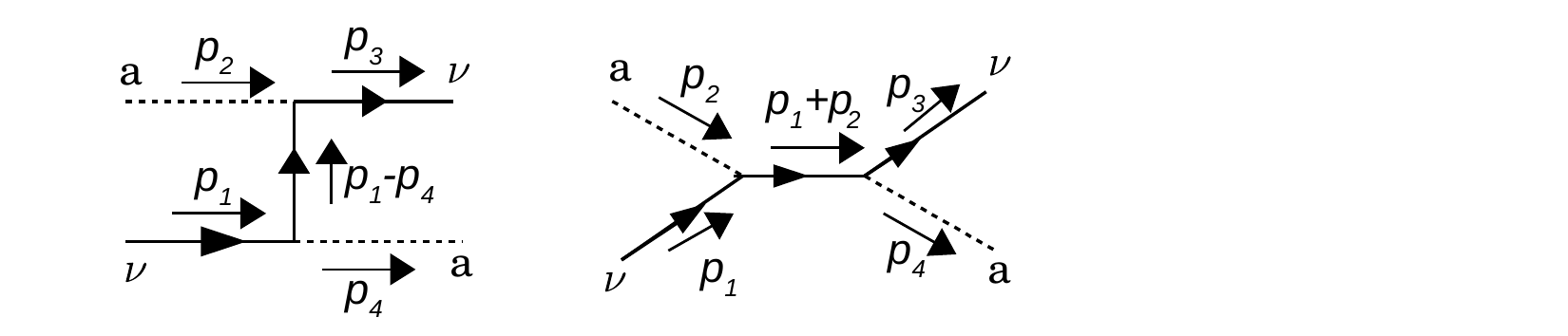}
\caption{Neutrino-UALP interaction channels considered.}
\label{fig1-diagrams}       
\end{figure}

The possible interaction channels for neutrinos and UALPs are shown in Fig. \ref{fig1-diagrams} and the corresponding scattering amplitudes 
for the left and right diagrams are
\be
\mathcal{M}_{\rm u}&=&-\frac{ig_i^2}{(p_1-p_4)^2-m_i^2}\bar{\nu}_i(p_3)[\slashed{p}_1-\slashed{p}_4+ m_i]\nu_i(p_1) \\
\mathcal{M}_{\rm s}&=&-\frac{ig_i^2}{(p_1+p_2)^2-m_i^2}\bar{\nu}_i(p_3)[\slashed{p}_1+\slashed{p}_2+m_i ]\nu_i(p_1). 
\ee

The total squared average amplitude is
\be
 \left|  \bar{\mathcal{M}}  \right|^2 = \frac{1}{2}\left(   \left|\mathcal{M}_{\rm u}\right|^2+ \mathcal{M}_{\rm u}\mathcal{M}_{\rm s}^*+ \mathcal{M}_{\rm s}\mathcal{M}_{\rm u}^*  +\left|\mathcal{M}_{\rm s}\right|^2   \right), 
\ee
where for the different terms we obtain the following expressions valid if $m_a \ll m_i$ and $E_3\lesssim E_1$:

\begin{equation}
 \left|\mathcal{M}_{\rm u}\right|^2=  \frac{g_i^4}{E_3} 
  \left[ {2 E_1}   + \frac{8 m_i^4}{E_3m_a^2} - \frac{8m_i^2}{ m_a} \right],  
\end{equation} 
\begin{equation}
 \left|\mathcal{M}_{\rm s}\right|^2=  \frac{g_i^4}{E_1}  
 \left[ 2 {E_3} + 8  \frac{m_i^2}{m_a} + 8 \frac{m_i^4}{E_1 m_a^2}  \right],
\end{equation}
and
\begin{equation}
 \left|\mathcal{M}_{\rm s}\mathcal{M}_{\rm u}^*\right|=  \left|\mathcal{M}_{\rm u}\mathcal{M}_{\rm s}^*\right|=4{g_i^4}  \left[2 \frac{(E_1 - E_3) }{m_a E_1 E_3} m_i^2 + 4 \frac{m_i^4}{E_1 E_3 m_a^2}-1\right]
\end{equation}

For the neutrino masses, we consider two usual assumptions for their values masses of the neutrinos \cite{pdg2020}: in the first one, the three masses are accommodated following a normal ordering (NO), i.e. $m_1\ll m_2< m_3$ with
\be
m_1&=& 10^{-5}{\rm eV} \\
m_2&=&8.7 \times 10^{-3}{\rm eV} \\
m_3&=&5 \times 10^{-2}{\rm eV},
\ee
 and the other option considered follows an inverted ordering (IO), $m_3\ll m_2< m_1$, with  \be
m_1&=& 4.92\times 10^{-2}{\rm eV} \\
m_2&=& 5 \times 10^{-2}{\rm eV} \\
m_3&=& 10^{-5}{\rm eV}.
\ee
{We note that the actual value of lightest neutrino mass is unknown, and we set it at $10^{-5}$ eV in order to produce our illustrative results.  }

The differential cross section in the laboratory (lab) frame is:

\be
\frac{d\sigma_{\nu a}}{dt_M}= \frac{1}{64 \pi s_M}\frac{1}{|\vec{p}_{1,\rm cm}|^2}|\mathcal{M}|^2, 
\ee
where $s_M=(p_1+p_2)^2$, and 
$
t_M= (p_3-p_1)^2
$
are the usual Mandelstam variables. The momenta of the initial and final neutrino are $p_1$ and $p_3$, respectively, while $p_2$ and $p_4$ refer to the initial and final UALP, respectively. The spatial momenta of the incident neutrino in the center of mass (CM) frame is $\vec{p}_{1,\rm cm}$, and it satisfies that $|\vec{p}_1|^2= E_{1,\rm cm}^2-m_i^2$, with
$$
E_{1,{\rm cm}}= \frac{s_M+m_i^2-m_a^2}{2\sqrt{s_M}}.
$$

\begin{figure*}
\includegraphics[width=0.99\textwidth,trim=5 20 45 0,clip]{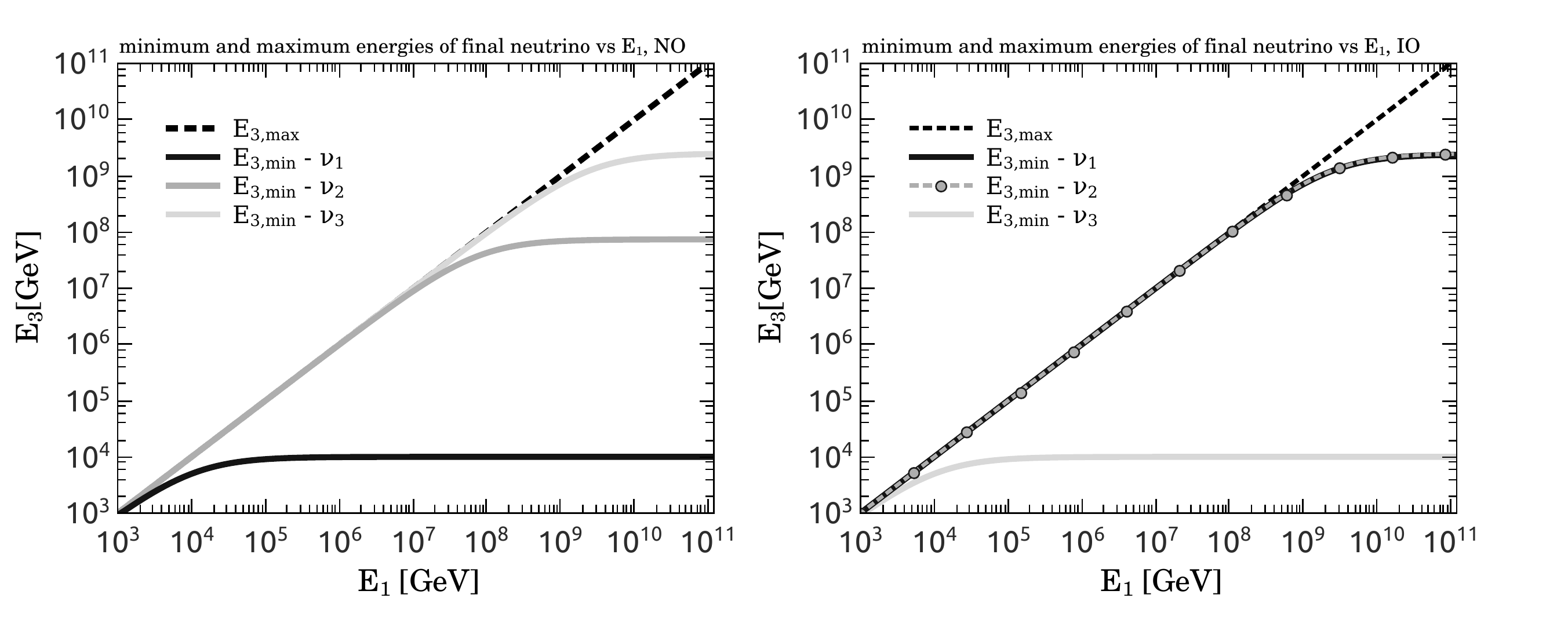}
\caption{Minimum and maximum energies of the final neutrino as a function of the initial neutrino energy $E_1$ for $\nu_1,\ \nu_2,$ and $\nu_3$.}
\label{fig2-E3minE3max}       
\end{figure*}

\begin{figure*}
\includegraphics[width=0.99\textwidth,trim=5 25 35 0,clip]{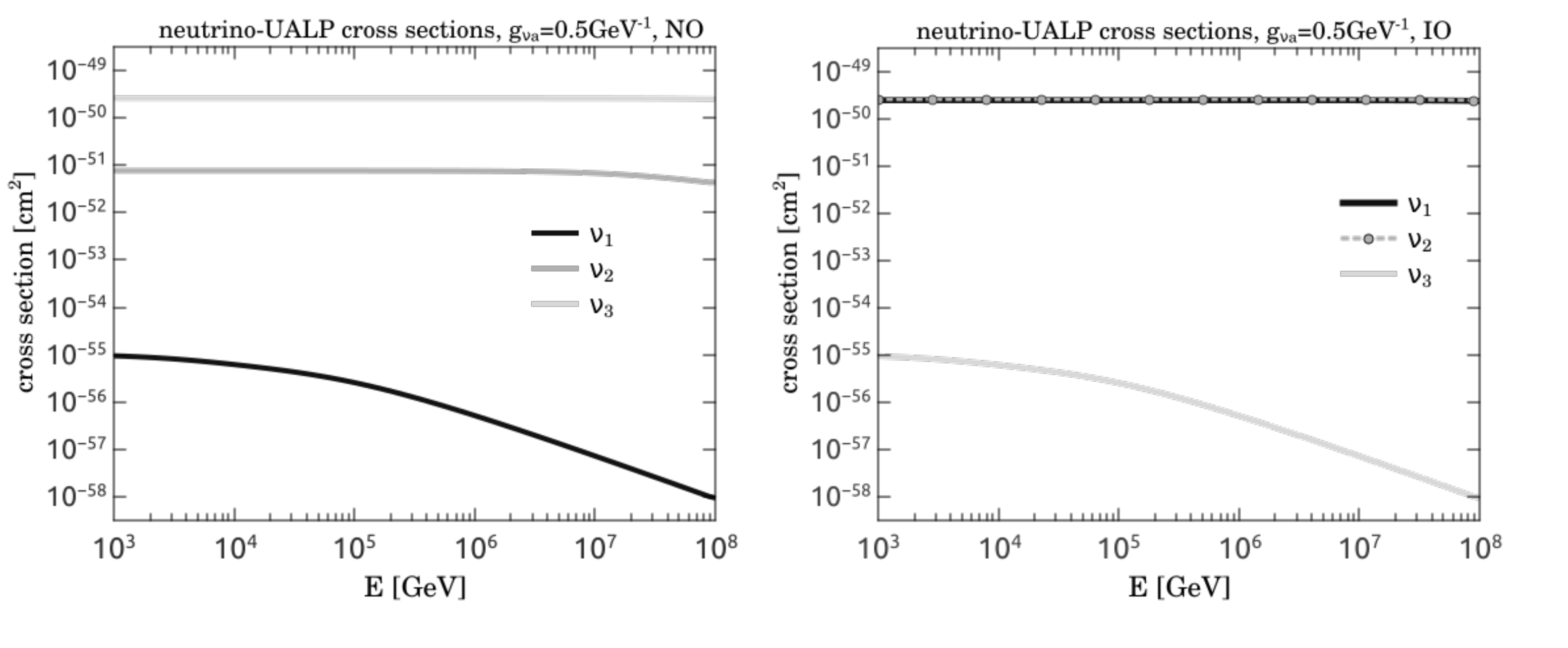}
\caption{Total cross sections as a function of the neutrino energy for the three massive neutrinos.}
\label{fig3-sigmas}       
\end{figure*}

In the lab frame, in turn, $t_M= -2m_a(E_1-E_3)$, and the differential cross section can be expressed as
\be
\frac{d\sigma_{\nu a}}{dE_3}= \frac{m_a}{32 \pi s_M}\frac{1}{|\vec{p}_{1,\rm cm}|^2}|\mathcal{M}|^2.  
\ee
The total cross section is, therefore,
\be
\sigma_{\nu a}=\int_{E_{3\rm min}}^{E_{3\rm max}}
\frac{d\sigma_{\nu a}}{dE_3},
\ee
where the minimum and maximum energies of the final neutrino in the lab frame are
\be
E_{3{\rm min}}&=& \gamma\left(E_{3\,{\rm cm}}- \beta\sqrt{E_{3{\rm cm}}^2-m_i^2}\right) \label{E3min}\\
E_{3{\rm max}}&=& \gamma\left(E_{3\,{\rm cm}}+ \beta\sqrt{E_{3{\rm cm}}^2-m_i^2}\right), \label{E3max}
\ee
with $\gamma=(E_1+m_a)/\sqrt{s_M}$, $\beta=(1-\gamma^2)^{-1/2}$, and $E_{3\rm cm}=E_{1\rm cm}$. The latter energy values are plotted in Fig. \ref{fig2-E3minE3max} as a function of the initial neutrino energy $E_1$. It can be seen that the energy loss per interaction is very small for $E_1\lesssim 10^8 {\rm GeV}$, except for the lightest neutrino, i.e., $\nu_1$ in the NO case, and $\nu_3$ in the IO case. 
The total cross sections are shown in Fig. \ref{fig3-sigmas} for $g=0.5 \,{\rm GeV}^{-1}$, and $m_a=5\times 10^{-22}{\rm eV}$, where the coupling for each massive neutrino is given by $g_i=2m_i g_{\nu a}$.

{In the above calculations, we have assumed for simplicity that the UALP is at rest, as it is expected to constitute non-relativistic DM (e.g. \citep{niemeyer2020}  ). Even in the galactic halo, considering that its dispersion velocity can be $\sim 100 {\rm km \ s^{-1}}$ \citep{hui2016}, this leads to a Lorentz factor as low as $\Gamma\simeq 1.0000005$ with respect to a reference frame fixed to the UALP. 
Since the strictly correct energies in the lab system should be multiplied by a factor $\approx\Gamma$, a relative error of $\epsilon_E\approx(\Gamma-1)$ is assumed, and this leads to an relative error in the kinematic variables $s_{\rm M}$ and $t_{\rm M}$ of $\epsilon_{s}\approx 2(\Gamma-1)$. Hence, a simple estimate of error propagation in the calculation of $d\sigma_{\nu a}/dE_{3}$ implies that the approximation of the UALP being at rest is accurate up to a negligible error much below $1\%$.} 

{ Throughout this work, we shall adopt coupling values $g_{i} \sim (0.1 - 1) {\rm GeV^{-1}}$, which are consistent with a cutoff for the effective field theory $\Lambda \sim (1-12) {\rm GeV}$. Still, the CM energies of the $\nu_i-a$ interactions, $\sqrt{s_M}=( m_i^2+m_a^2+ 2m_a E_{1})^{0.5}$, are well below this cutoff for a UALP mass $m_a=5\times 10^{-22}{\rm eV}$ and neutrino energies $E_1< 10^8{\rm GeV}$. Therefore, the validity of the effective operator approach is not compromised in the present context.}

{}

\section{Neutrino propagation including interaction effects}\label{sec:3propagation}
In this section we describe the propagation of a diffuse flux of high energy neutrinos of astrophysical origin, considering separately the extragalactic propagation and then the propagation through the DM halo of our galaxy. 

\subsection{Injection by astrophysical sources} 
The neutrino injection depends on the proton acceleration mechanism operating at the astrophysical sources, and in particular, we are interested in the corresponding initial flavor ratios $f_{\alpha,\rm s}(E)$, i.e., at the beginning of the propagation stage. If the production channel is neutron decay ($n\rightarrow p \ \bar{\nu}_e e^-$), then only the electron flavor is generated, and hence the flavor composition at the sources is simply $(f_{e,\rm s}:f_{\mu,\rm s}:f_{\tau,\rm s})=(1:0:0)$. 
However, neutrino production after pion decays is more promisory as pions can be efficiently created by $pp$ and $p\gamma$ interactions at the sources\footnote{Neutrons are also efficiently created in these interactions, but the generated neutrino carries only a very small fraction of the partent neutron energy ($E_\nu/E_n\sim 10^{-3}$) \cite{lipari2007}}. 

We then consider two additional source benchmarks: in the first one, both pions and muons decay unaffected by synchrotron losses leading to the usually expected flavor composition at the sources $(f_{e,\rm s}:f_{\mu,\rm s}:f_{\tau,\rm s})=(1:2:0)$, and we reffer to this case as pion decay sources. In the other typical case, muons are completely depleted due to synchrotron losses and we have a proportion of flavors $(f_{e,\rm s}:f_{\mu,\rm s}:f_{\tau,\rm s})=(0:1:0)$ at these highly magnetized muon-damped sources. There are also other intermediate possibilities for which muons could be cooled significantly only above a certain energy, and in these cases the flavor at the sources would be energy dependent \cite{icecubegen2,bustamantemagneto2020}. Still, similarly to the assumed in previous works \cite{mehta2011,icecube2015,palatable2015,inferring2019,palladino2019}, here we consider the mentioned three benchmark cases of initial flavor compositions, as our main goal is to focus on the effects of the possible neutrino interactions with UALP DM.
{To achieve this, we adopt for the neutrino injection} a canonical power-law dependence with the neutrino energy times an exponential cutoff at $E_{\rm c}= 10^{7}{\rm GeV}$:
\be
Q_{\nu_i}(E,z)= K_\nu \frac{\psi_{\rm SFR}(z)}{\psi_{\rm SFR}(z=1)} f_{i,\rm s} E^{-\alpha}\exp\left(-\frac{E}{E_{\rm c}}\right).
\ee
Here, $K_\nu$ is a normalization constant, and the factors $f_{i,\rm s}$ depend on the flavor composition at the sources since the neutrino injection corresponding to the flavor $\alpha$ is given by
\be 
Q_{\nu_\alpha}(E,z)=\sum_{i=1}^{3}\left|U_{\alpha i}\right|^2 f_{i,\rm s}Q_{\nu_i}(E,z).
\ee
The evolution with the redshift $z$ is assumed to follow the star formation rate (SFR) as given by \cite{madau2014},
\be
\psi_{\rm SFR}(z)= 0.015\frac{(1+z)^{2.7}}{1+ \left[(1+z)/2.9\right]^{5.6}} \ M_{\odot}{\rm yr^{-1}Mpc^{-3} }.
\ee
{Here we remark that we assume that the sources are also isotropically distributed in the sky, and the approach we apply in this work is adequate to capture the possible modification the total diffuse neutrino flux by accounting for the energy loss undergone by neutrinos if they interact with UALP DM. A study of these effects in the case of individual point sources should also account for the angular deflection generated by the scatterings, as it was done in Ref. \citep{choi2019} for the flare neutrino event IceCube-170922A associated to the blazar TXS 0506+056 \citep{IceCube170922A}.} 

In the present context, and for illustration of the effects of the described interactions on a diffuse neutrino flux, we choose to normalize the injection as 
\begin{multline}
\left.\frac{c}{4\pi}\int_0^{5}\frac{dz}{(1+z)^2} \frac{dt}{dz} Q_{\nu_\mu}\left(E(1+z),z\right)\right|_{E=10^5{\rm GeV}} \\= \left. \Phi^{\rm IceCube}_{\nu_\mu+\bar{\nu}_\mu}(E)\right|_{E=10^5{\rm GeV}},
\end{multline}
so that the $\nu_\mu+\bar{\nu}_\mu$ flux obtained in the absence of $\nu-a$ interactions matches the best fit flux obtained by IceCube evaluated at $10^5$GeV \cite{stettner2019}, $$\Phi^{\rm IceCube}_{\nu_\mu+\bar{\nu}_\mu}(E)=\Phi_0\ \left(\frac{E}{10^5 {\rm GeV}}\right)^{-2.28},$$
where $\Phi_0=1.44\times 10^{-18}{\rm GeV^{-1}s^{-1}sr^{-1}cm^{-2}}$, and we accordingly assume $\alpha=2.28$ for the index of injection.

 \subsection{Extragalactic propagation} 

We treat the extragalactic propagation of astrophysical neutrinos making use of the following transport equation with continuous losses for the comoving density of each massive neutrino \cite{venya2006,farzan2014}
 \begin{multline}
\frac{\partial N_{\nu_i}(E,t)}{\partial t}= Q_{\nu_i}(E,t)-3H(t)N_{\nu_i}(E,t)\\+ \frac{\partial\left[H(t)E\ N_{\nu_i}(E,t)+N_{\nu_i}(E,t)b_{\nu a}(E,t)\right]}{\partial E},
 \end{multline}
which accounts for the effect of expansion of the universe through the terms proportional to 
$$H(t)= H_0\left[\Omega_{\rm m}(1+z)^3+\Omega_\Lambda\right]^{-1},$$
 and the interactions with DM UALPs are characterized by the corresponding continuous energy loss rate \cite{alhersanchordoqui2010}:
\be
b_{\nu a}(E,z)= n_{\rm dm}c\int_{E_{3\rm min}}^{E_{3\rm max}} dE_3 (E-E_3) \frac{d\sigma_{\nu a}(E,E_3)}{dE}.\label{bnua}
\ee 
Here, the density of extragalactic DM is given by 
\be
n_{\rm dm}(z)= \left(\frac{3H_0^2}{8\pi G}\right)\frac{\Omega_{\rm m}}{m_a}(1+z)^3, 
\ee
with $H_0\simeq 70\,{\rm km \, s^{-1}}$ and $\Omega_{\rm m}\simeq 0.3$. For instance, the extragalactic DM column $X_{\rm dm}(z)=\int_0^z n_{\rm dm}(z') dz'$ for sources at $z\approx 1$ is  $X_{\rm dm}(1) \approx 6\times 10^{52}{\rm cm^{-2}} $, and for sources at a redshift $z\approx 5$ it is $X_{\rm dm}(5) \approx 6\times 10^{53}{\rm cm^{-2}} $. {In Eq. (\ref{bnua}), $E$ represents the initial neutrino energy and we integrate over the final neutrino energy $E_3$ in a range that can be seen in Fig. \ref{fig2-E3minE3max} and is determined by $E_{3\rm min}$ and $E_{3\rm max}$ as defined above above in Eqs.(\ref{E3min},\ref{E3max}). } 
We note that in the NO case for $\nu_2$ and $\nu_3$, the energy loss per interaction is very small for $E_1\lesssim 10^8$GeV, and hence the continuous loss approximation is justified. However, this is not the case for $\nu_1$, given that the range of initial neutrino energies $E_1$ for a fixed final energy $E_3$ is much broader, and additionally, the differential cross section $\frac{d\sigma_{\nu a}}{dE_3}$ is actually peaked at $E_1\ll E_3$. Therefore, the present approach is accurate to obtain the neutrino fluxes for a coupling 
\be
g_{\nu a}\lesssim 1.5 {\rm GeV}^{-1} 
\ee
such that no significant $\nu_1-a$ interactions take place, i.e.,
\be 
\sigma_{\nu a}&<& \frac{1}{X_{\rm dm}(z=5)}\approx 1.6\times 10^{-54}{\rm cm^{2}}. 
\ee 
A similar reasoning can be applied to the IO case, where the heavier neutrinos are $\nu_1$ and $\nu_2$, and the lightest one will be affected only for the mentioned range of large couplings. 
For higher values of $g_{\nu a}$, the only surviving flux would be the one of $\nu_1$ in the NO case and the one of $\nu_3$ in the IO case, so that the corresponding flavor composition would be clearly determined as 
\be (f_e:f_\mu:f_\tau)_{\rm NO}&=&|(U_{e1}|^2:|U_{\mu 1}|^2:|U_{\tau 1}|^2) \\
(f_e:f_\mu:f_\tau)_{\rm IO}&=&|(U_{e3}|^2:|U_{\mu 3}|^2:|U_{\tau 3}|^2),
 \ee 
 if absorption is not complete, but affects only the two heavier neutrinos.
  More on this issue is discussed below.

In terms of redshift, considering that $dt/dz=-H(z)(1+z)$, we can write the transport equation as
\begin{multline}
  \frac{\partial N_{\nu_i}(E,z)}{\partial z}=-\frac{Q_{\nu_i}(E,z)}{H(z)(1+z)} \\ +\left[\frac{1}{1+z}\right]\left[{2}- \frac{1}{H(z)}\frac{\partial b_{\nu  a}(E,z)}{\partial E}\right]N_{\nu_i}(E,z) \\ -\left[\frac{E}{1+z}+\frac{b_{\nu  a}}{H(z)(1+z)}\right]\frac{\partial N_{\nu_i}(E,z)}{\partial E}, \label{transportEqz}
\end{multline}
and the neutrino flux after the propagation is found as 
\be
{\Phi_{\nu_\alpha}}(E)=\frac{c}{4\pi}\sum_{i=1}^3f_{i,}{\left|U_{\alpha\,i}\right|^2N_{\nu_i} (E)}. \label{Eq.fluxafterpropagation}
\ee

We solve Eq. (\ref{transportEqz}) using the method of the characteristics, i.e., with the characteristic curves satisfying
\be
\frac{dE}{dz}= \frac{E}{1+z}+\frac{b_{\nu a}}{H(z)(1+z)}, 
\ee
we obtain the solutions
\begin{multline} 
N_{\nu_i}(z,E)= \int_z^{z_{\rm max}}dz'\frac{Q_{\nu_i}(z',E')}{H(z')(1+z')}
 \\ \times \exp{\left[-\int_z^{z'}
\frac{dz''}{1+z''}\left(2-\frac{1}{H(z'')}\frac{\partial b_{\nu a}}{\partial E}\right) \right]},
\end{multline}
where we assume $z_{\rm max}=5$. The results obtained for $z=0$ are then used as input to describe the propagation through our galaxy, as discussed below.

\begin{figure}
\includegraphics[width=0.5\textwidth,trim=25 185 20 20,clip]{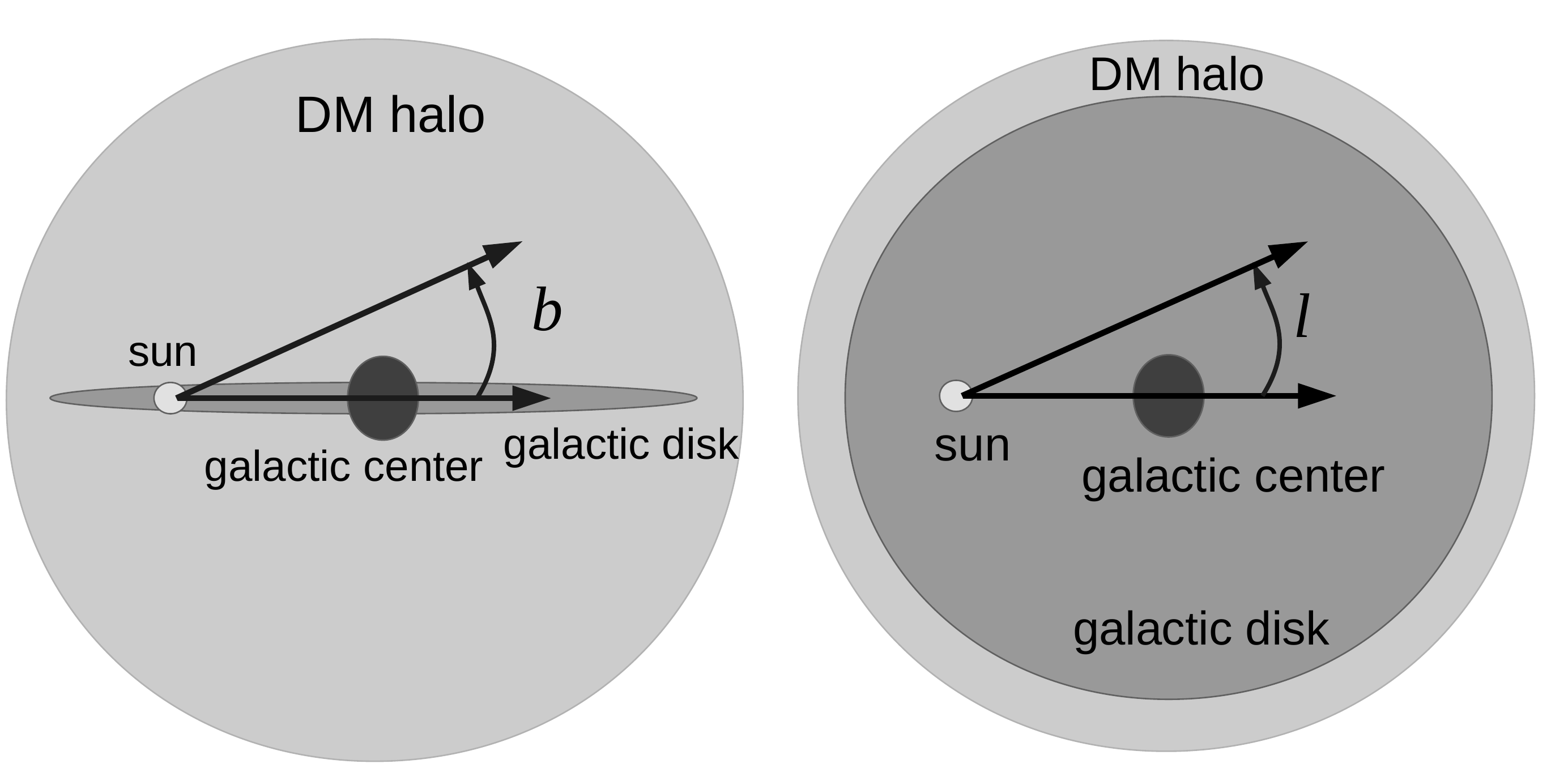}
\caption{Side and top schematic views of the galactic DM halo, indicating the galactic coordinates $l$ and $b$.}
\label{fig4-halo}       
\end{figure}

\subsection{Propagation through the DM halo} 
After propagation outside our galaxy, a diffuse flux of neutrinos arriving to the Milky Way DM halo has to traverse it at different directions, facing different DM column depths $X_{\rm dm,h}(l,b)$ depending on the galactic longitudes $l$ and latitudes $b$ (see Fig. \ref{fig4-halo}). The DM density profile adopted is a generalized spherically symmetric Navarro, Frenk, and White (NFW) one \cite{nfw,benito2019},
\be
n_{\rm dm,h}(r)= \frac{\rho_{\rm s}}{m_a}\left(\frac{r}{r_{\rm s}}\right)^{-\gamma_{0}}\left(1+\frac{r}{r_{\rm s}}\right)^{\gamma_0-3}, 
\ee
where $\gamma_0=1.2$, $r$ is the distance to the galactic center, $r_{\rm s}=20\,{\rm kpc}$, and the density there ($\rho_{\rm s}$) is obtained by assuming a local density $\rho_0=0.4 \, {\rm GeV \, cm^{-3}}$ at the position of our solar system $r_0=8.127 \,{\rm kpc}$.
For instance, we obtain $X_{\rm dm,h}(l=0,b=10^\circ)= 2\times 10^{53}{\rm cm^{-2}}$ and $X_{\rm dm,h}(l=0,b=90^\circ)= 3\times 10^{52}{\rm cm^{-2}}$. The transport equation in the absence of injection and adiabatic losses can be written as 
\be
\frac{\partial N_{\nu_i}(E,l,b)}{\partial X}= \frac{1}{ n_{\rm dm,h} c }\frac{\partial \left[ b_{\nu a,\rm h}(E) N_{\nu i}(E,l,b)\right] }{\partial E},\label{transportEqhalo} 
\ee
where $b_{\nu a,\rm h}(E)$ is analogous to Eq.(\ref{bnua}) but with the DM density corresponding to the galactic halo. 

Using again the method of the characteristics, we find that the solution of Eq.(\ref{transportEqhalo}) can be expressed as
\be 
N_{\nu_i}(E,l,b)= N_{\nu_i}(E,z=0) \left[\frac{b_{\nu a}\left(E'({X'=0})\right) }{b_{\nu a}(E)}\right], 
\ee
where $E'$ is the characteristic energy corresponding to a depth $X'$ such that the final energy for a depth $X_{\rm dm,h}(l,b)$ is $E$.


\section{Results}\label{sec4:results}

\begin{figure*}
\includegraphics[width=\textwidth,trim=0 55 60 0,clip]{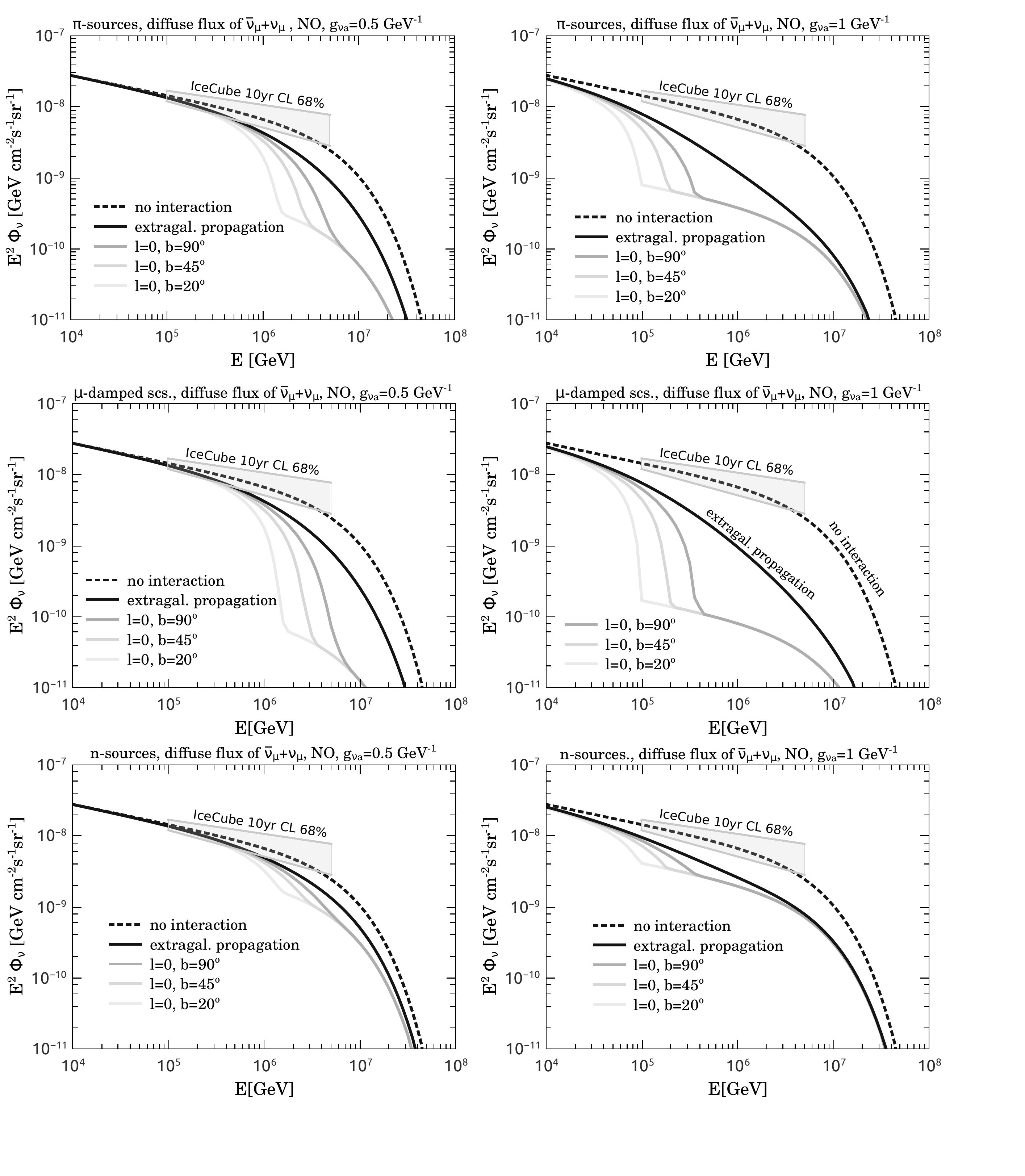} 
\caption{Diffuse fluxes of $\nu_\mu+\bar{\nu}_\mu$ assuming a normal ordering for the neutrino masses, for $g_{\nu a}=0.5\,{\rm GeV^{-1}}$ and $g_{\nu a}=1\,{\rm GeV^{-1}}$ in the left and right panels, respectively. Top panels correspond to pion decay sources, middle panels to muon damped sources, and bottom panels to neutron decay sources. We show the fluxes corresponding to the case of no interactions in dashed lines, and for reference we include the 10-year CL 68\% IceCube  data corresponding to muon tracks. The flux corresponding to interactions with extragalactic DM only is indicated with a black solid line in each panel, and the fluxes for the arrival directions given by $l=0^\circ$ and $b=\left\{90^\circ, 45^\circ,0^\circ\right\}$ are marked with curves in dark gray, gray, and light gray, respectively.    }
\label{fig5-numufluxes}       
\end{figure*}

\begin{figure*}
\includegraphics[width=\textwidth,trim=0 55 60 0,clip]{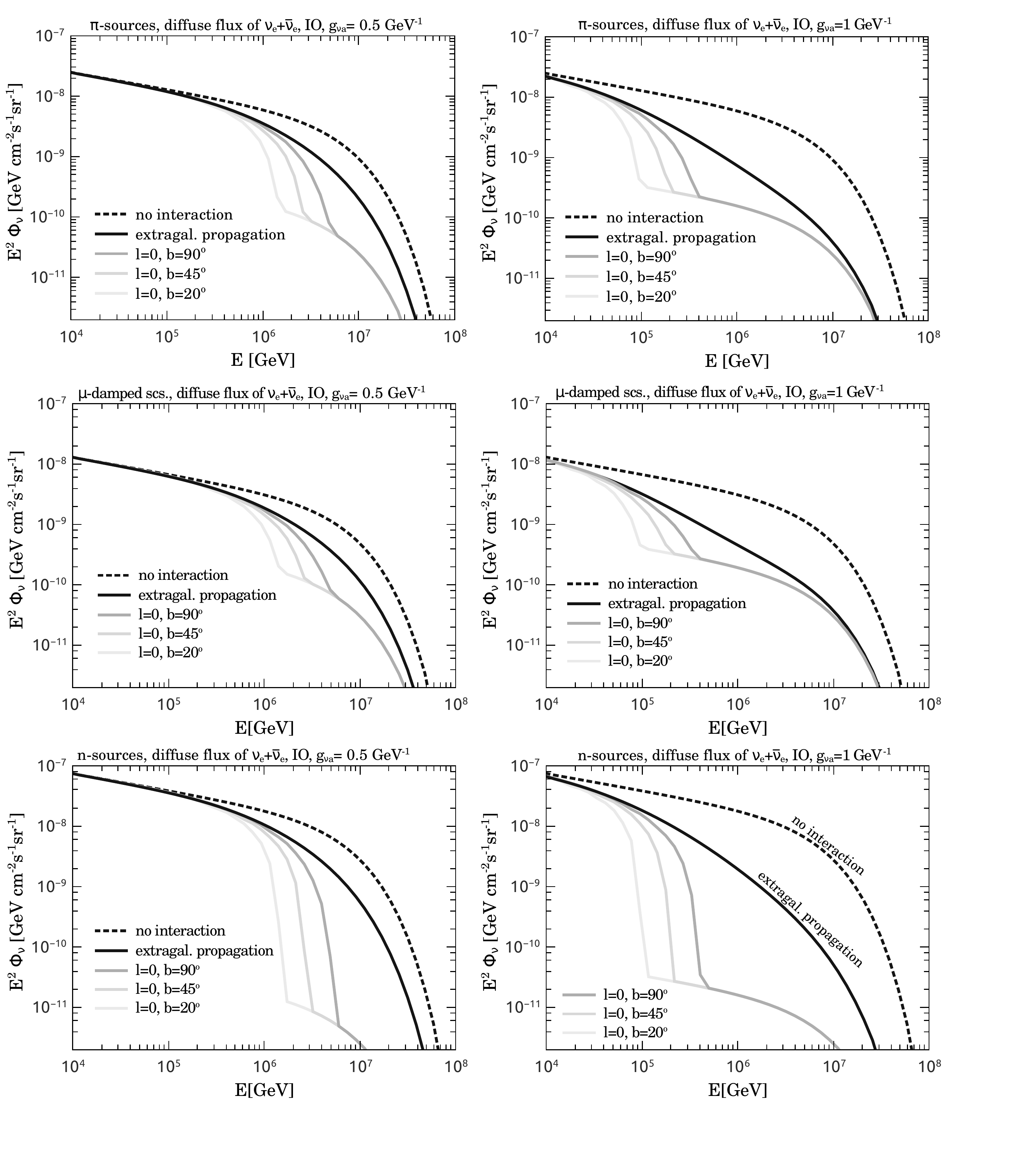} 
\caption{Diffuse fluxes of $\nu_e+\bar{\nu}_e$ assuming an inverted order for the neutrino masses, for $g_{\nu a}=0.5\,{\rm GeV^{-1}}$ and $g_{\nu a}=1\,{\rm GeV^{-1}}$ in the left and right panels, respectively. Top panels correspond to pion decay sources, middle panels to muon damped sources, and bottom panels to neutron decay sources. The flux corresponding to interactions with extragalactic DM only is indicated with a black solid line in each panel, and the fluxes for the arrival directions given by $l=0^\circ$ and $b=\left\{90^\circ, 45^\circ,0^\circ\right\}$ are marked with curves in dark gray, gray, and light gray, respectively.   }
\label{fig6-nuefluxes}       
\end{figure*}

In this section, we present the obtained results for the neutrino fluxes after extragalactic and galactic propagation including the effects of $\nu-a$ scatterings.
As discussed above, we consider three different benchmark possibilities for the flavor composition at the sources.

The proportion of the different massive neutrinos emitted at the sources $(f_{1,\rm s}:f_{2, \rm s}:f_{3,\rm s})$ in the different cases assumed is such that
\be 
f_{i,\rm s}\propto \left\lbrace \begin{array}{c c}
\left(|U_{ei}|^2+2|U_{\mu i}|^2\right) & \ \mbox{for $\pi$ and $\mu$ decays} \\
\left(|U_{\mu i}|^2\right) & \ \mbox{for $\mu$ damped case} \\
\left(|U_{ei}|^2\right) & \ \mbox{for $n$ decays} 
\end{array} \right. \nonumber .
\ee
 
The neutrino propagation including  the effect of interactions is described by the solution of the trasport equations for each massive neutrino as mentioned above, and the flux of neutrinos of the different flavors is computed by superposition as in Eq. (\ref{Eq.fluxafterpropagation}). In Fig. \ref{fig5-numufluxes}, we show the obtained fluxes of $\nu_\mu+ \bar{\nu}_\mu$ in the NO case including the effects of interactions for $g_{\nu a}=0.5 {\rm GeV^{-1}}$ and $g_{\nu a}=1 {\rm GeV^{-1}}$. We show the fluxes after extragalactic propagation only as well as the arriving in directions given by $l=0^\circ$ and $b=(20^\circ, 45^\circ, 90^\circ)$, and we compare these results with the flux corresponding to no interactions and with the flux of $\nu_\mu+ \bar{\nu}_\mu$ measured by IceCube at CL 68\% based on 10 years of muon track data \cite{stettner2019}. It can be seen that the interaction effects  are more noticeable if the arrival directions are closer to the galactic center (GC) since the DM column along the neutrino path is greater. The effect of increasing the coupling strength can be appreciated by comparing the results of the left panels to those of the right panels of Fig. \ref{fig5-numufluxes}: a more significant flux attenuation takes place as the coupling is increased, as can be expected due to a larger cross section. It can also be seen that the fluxes are most affected in the case of muon damped sources and least affected for neutron decay sources. This can be understood taking into account that the cross section is higher for neutrino $\nu_3$, followed by that for $\nu_2$ and the weakest one is for $\nu_1$, as can be seen in Fig. \ref{fig3-sigmas}. Therefore, since in the case of muon-damped sources only one muon neutrino is produced, the fact that $|U_{\mu 3}|\approx 0.55$ implies that a significant portion of the emitted flux would be more affected by interactions. Likewise, in the neutron-decay case, only the electron flavor is produced and since $|U_{e 3}|\approx 0.022$, the net attenuation is weaker and due basically to the depletion corresponding to the neutrino $\nu_2$, while the $\nu_1$ remains unaffected for the coupling values considered.  

\begin{figure*}
\includegraphics[width=\textwidth,trim=0 70 430.2 0,clip]{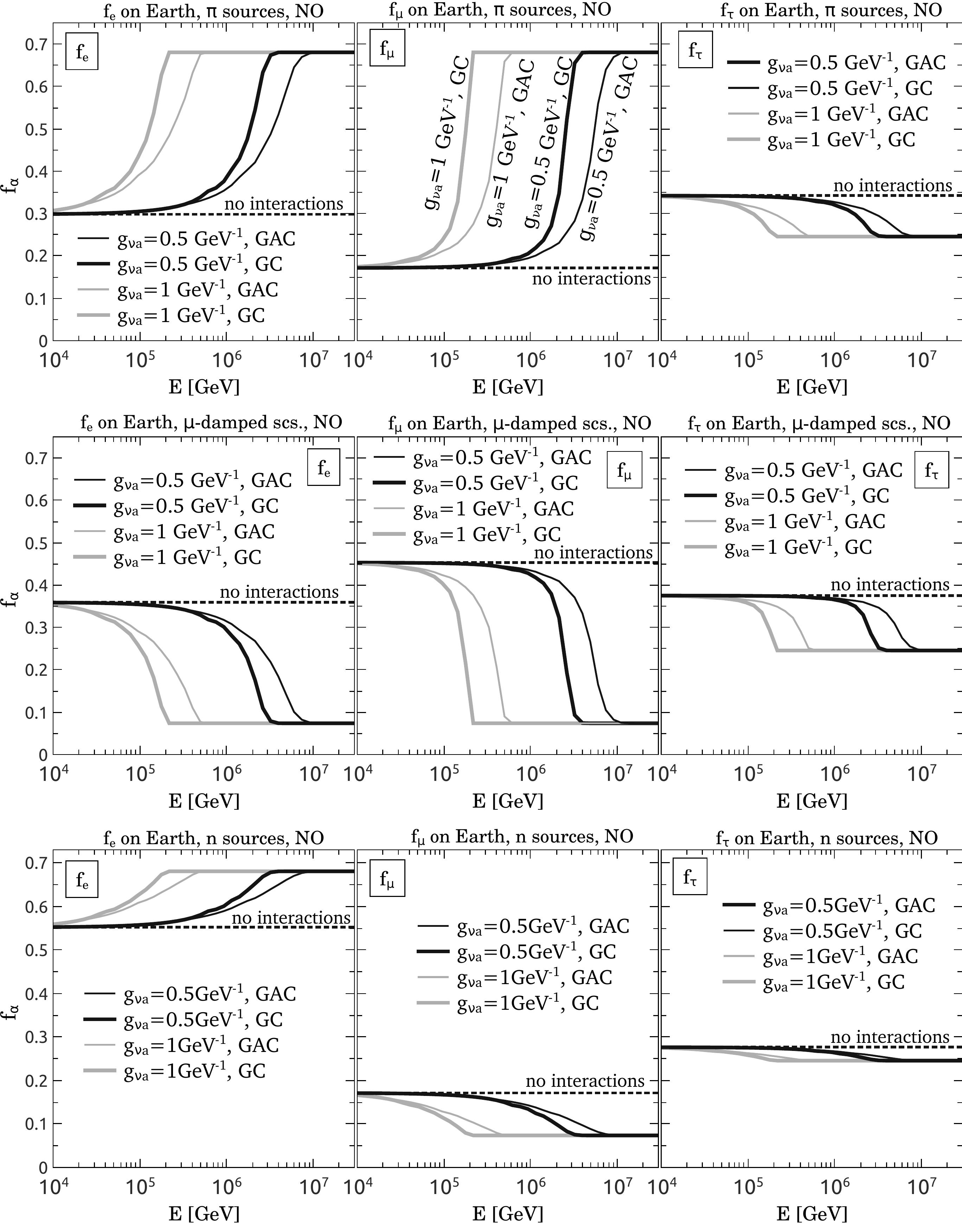}
\caption{Energy dependent flavor ratios of neutrinos reaching the Earth in the NO case {for the electron, muon, and tau flavors in the left, center, and right panels, respectively}. Thick solid lines correspond to the half of the sky centered around the galactic center (GC), while thin lines correspond to the other half of the sky containing the galactic anti-center (GAC). The results for no interactions are marked using dashed lines. Top panels correspond to pion decay sources, middle panels to muon damped sources, and bottom panels to neutron decay sources.}
\label{fig7-flavor-ratios-E-NO}       
\end{figure*}

\begin{figure*}
\includegraphics[width=\textwidth,trim=0 70 430.5 0,clip,clip]{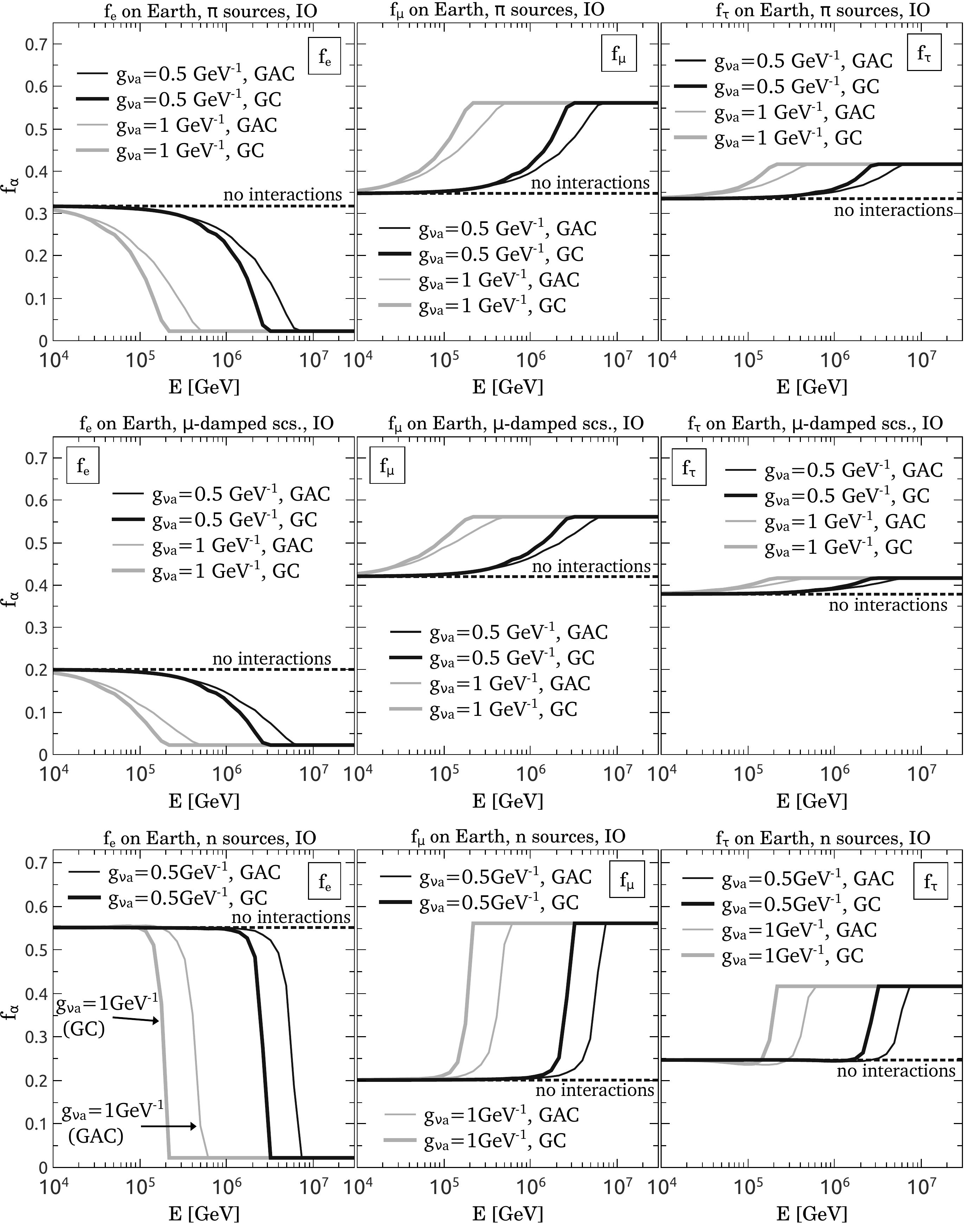}
\caption{Energy dependent flavor ratios of neutrinos reaching the Earth in the IO case {for the electron, muon, and tau flavors in the left, center, and right panels, respectively}. Thick solid lines correspond to the half of the sky centered around the galactic center (GC), while thin lines correspond to the other half of the sky containing the galactic anti-center (GAC). The results for no interactions are marked using dashed lines. Top panels correspond to pion decay sources, middle panels to muon damped sources, and bottom panels to neutron decay sources.}
\label{fig8-flavor-ratios-E-IO}       
\end{figure*}

In Fig. \ref{fig6-nuefluxes}, we show the diffuse fluxes of the electron flavor neutrinos in the NO case, since in this flavor the fluxes are more affected. This is because in the IO case, the neutrino $\nu_3$ is the least affected one and its contribution to the electron flavor is very small, given that $|U_{e3}|\approx 0.022$ as in the NO case. In order to appreciate the contribution of the three flavors, we show the energy dependence of the flavor ratios on the Earth in Fig. \ref{fig7-flavor-ratios-E-NO} for the NO case and in Fig. \ref{fig8-flavor-ratios-E-IO} for the IO case. In these plots we include the averaged  results over two regions in the sky: one for $(-90^\circ<l<90^\circ)$, i.e., centered around the GC, and the other for $(90^\circ< l <270^\circ)$, i.e., centered around the galactic anti-center (GAC). Clearly, neutrinos corresponding to the former hemisphere face a higher column of galactic DM than those arriving corresponding to the GAC hemisphere, and attenuation is more significant. By comparing the results of Fig. \ref{fig7-flavor-ratios-E-NO} with the ones of Fig. \ref{fig8-flavor-ratios-E-IO}, it can be seen that the effects of interactions lead to very different flavor ratios if neutrino masses follow a NO or if they follow an IO. In the former case and in relation to the flavor composition expected without interactions, the electron flavor is increased and becomes dominant, followed by the tau flavor which is decreased, and the muon flavor contribution is significantly reduced. Quite the opposite is expected in the IO case if interactions are important: the muon flavor becomes dominant, followed by the tau flavor still with reduced relevance, and the electron flavor contribution becomes very small. 


\begin{figure*}
\includegraphics[width=\textwidth,trim=0 70 430 0,clip]{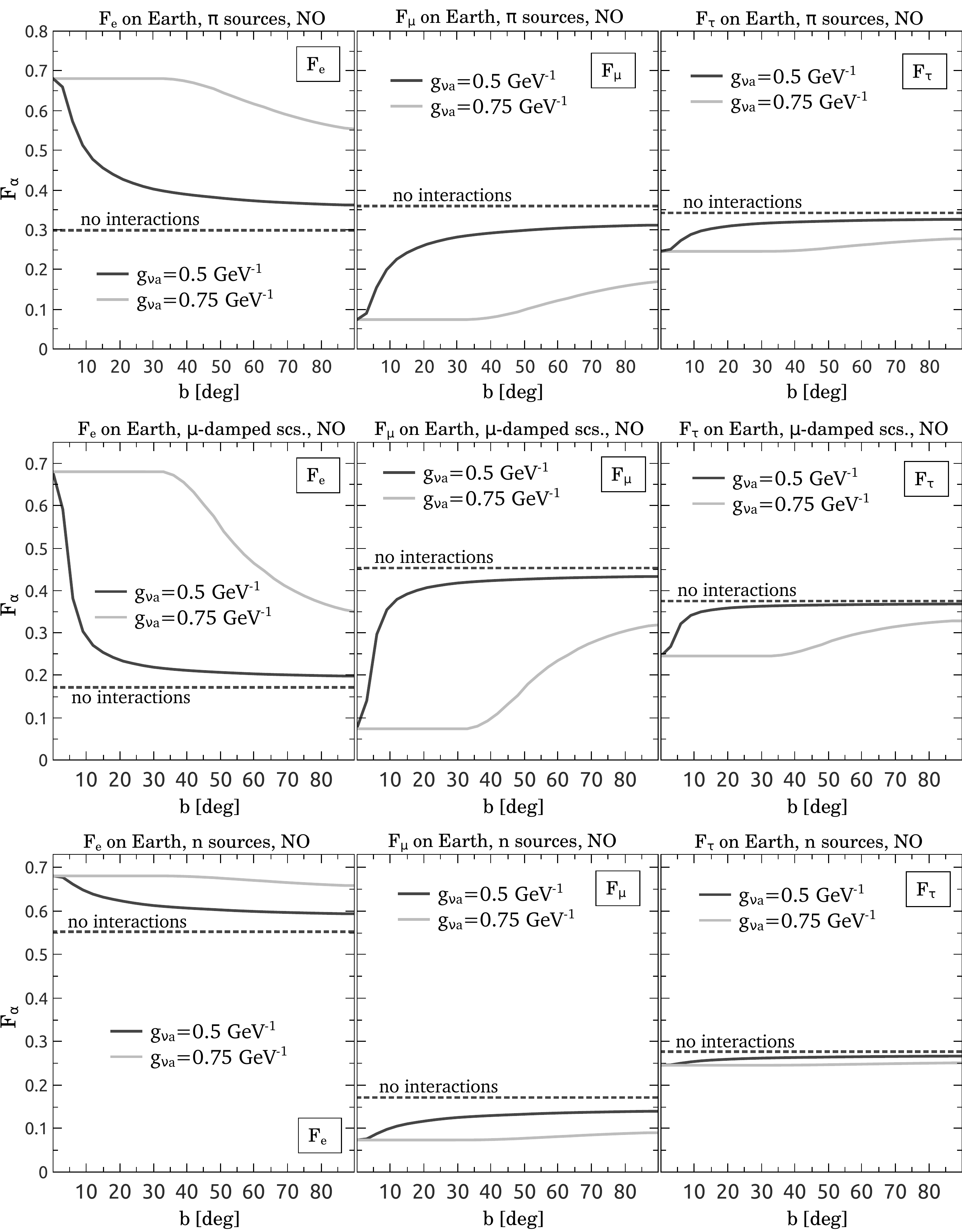}
\caption{{Flavor ratios of the integrated fluxes} in the NO case as a function of the galactic latitude $b$ corresponding to the arrival direction, for a fixed galactic longitude $l=0^\circ$. The electron, muon, and tau flavor ratios are shown in the left, middle, and right panels, respectively. Top panels correspond to pion decay sources, middle panels to muon damped sources, and bottom panels to neutron decay sources. }
\label{fig9-integrated-flavor-ratios-b-NO}       
\end{figure*}

\begin{figure*}
\includegraphics[width=\textwidth,trim=0 70 430 0,clip]{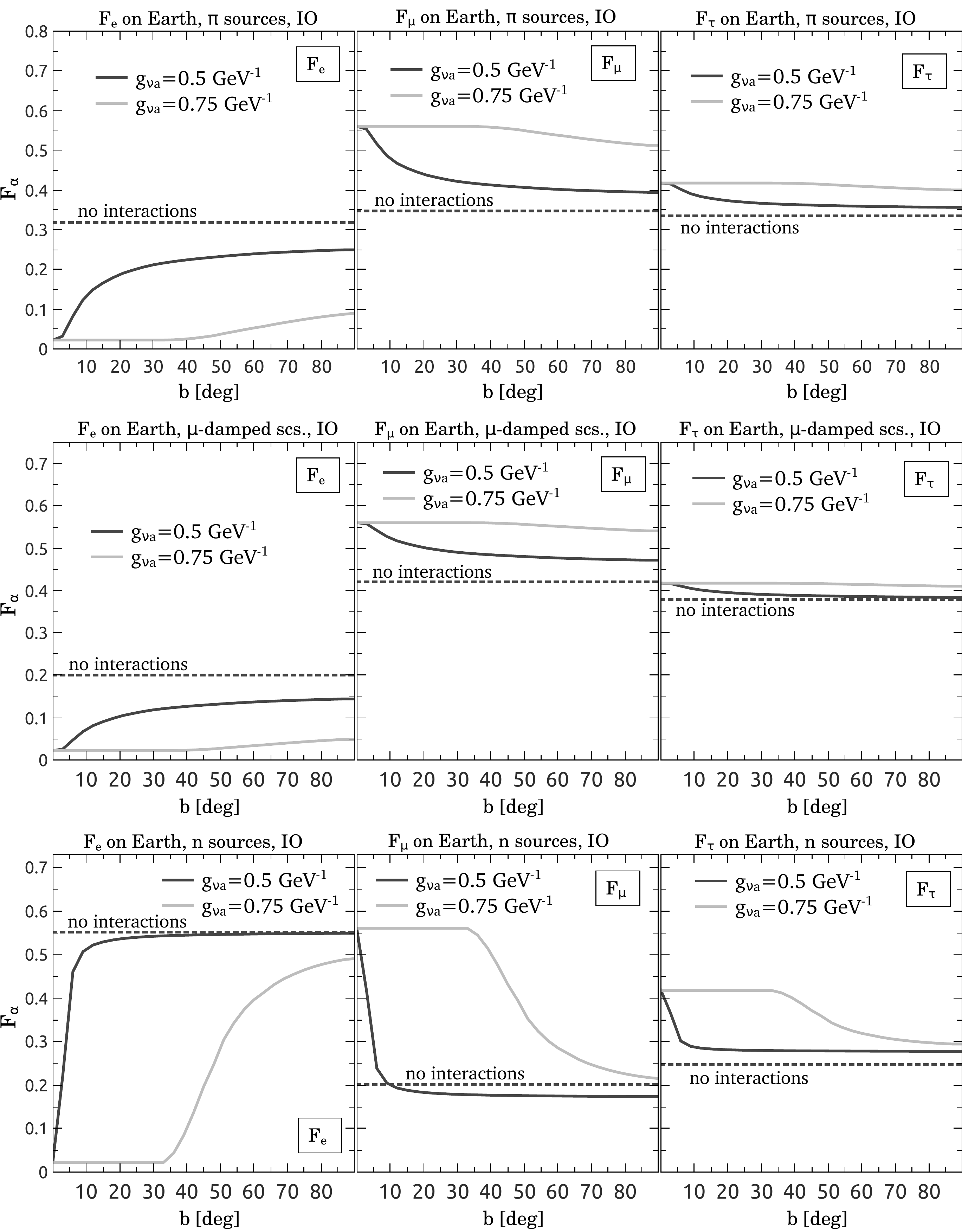}
\caption{{Flavor ratios of the integrated fluxes} in the IO case as a function of the galactic latitude $b$ corresponding to the arrival direction, for a fixed galactic longitude $l=0^\circ$. The electron, muon, and tau flavor ratios are shown in the left, middle, and right panels, respectively. Top panels correspond to pion decay sources, middle panels to muon damped sources, and bottom panels to neutron decay sources.}
\label{fig10-integrated-flavor-ratios-b-IO}       
\end{figure*}

{We can integrate on the neutrino energy the differential fluxes in order to obtain the flavor ratios of the integrated fluxes above $E_{\rm min}$ as:}
\be 
F_{\alpha}=\frac{\int_{E_{\rm min}}^\infty dE_\nu \Phi_{\nu_\alpha}(E_\nu)}
                     {\int_{E_{\rm min}}^\infty dE_\nu \left[\Phi_{\nu_e}(E_\nu)+ \Phi_{\nu_\mu}(E_\nu)+ \Phi_{\nu_\tau}(E_\nu)\right]}.
\ee
 The results for $E_{\rm min}=5\times 10^5{\rm GeV}$ are shown in Fig. \ref{fig9-integrated-flavor-ratios-b-NO}  as a function of the galactic latitude $b$ for the NO case and in Fig. \ref{fig10-integrated-flavor-ratios-b-IO} for the IO case. In this figures, we show the flavor ratios of the integrated fluxes for the three cases of sources assumed and with different values of the coupling, $g_{\nu a}=0.5\,{\rm GeV}^{-1}$ and $g_{\nu a}= 0.75\,{\rm GeV}^{-1}$. 
Also, as mentioned in Section \ref{sec:2interactions}, for high values of the coupling, only the $\nu_{1(3)}$  would survive in the NO(IO) case, and can be completely unaffected due to a much lower cross section. In these cases, the flavor ratios could also be homogeneous across the full sky, but fixed at the particular values 
\be 
  (F_e:F_\mu:F_\tau)_{\rm NO}&=& \left(|U_{e1}|^2:|U_{\mu 1}|^2:|U_{\tau 1}|^2\right) \\ 
   &\approx& (0.68: \, 0.075:  \, 0.24)\nonumber \\
  (F_e:F_\mu:F_\tau)_{\rm IO}&=& \left(|U_{e3}|^2:|U_{\mu 3}|^2:|U_{\tau 3}|^2\right) \\ 
   &\approx& (0.022: \, 0.56:  \, 0.41).\nonumber
\ee

\begin{figure*}
\begin{center}
\includegraphics[width=0.99\textwidth,trim=0 0 0 400,clip]{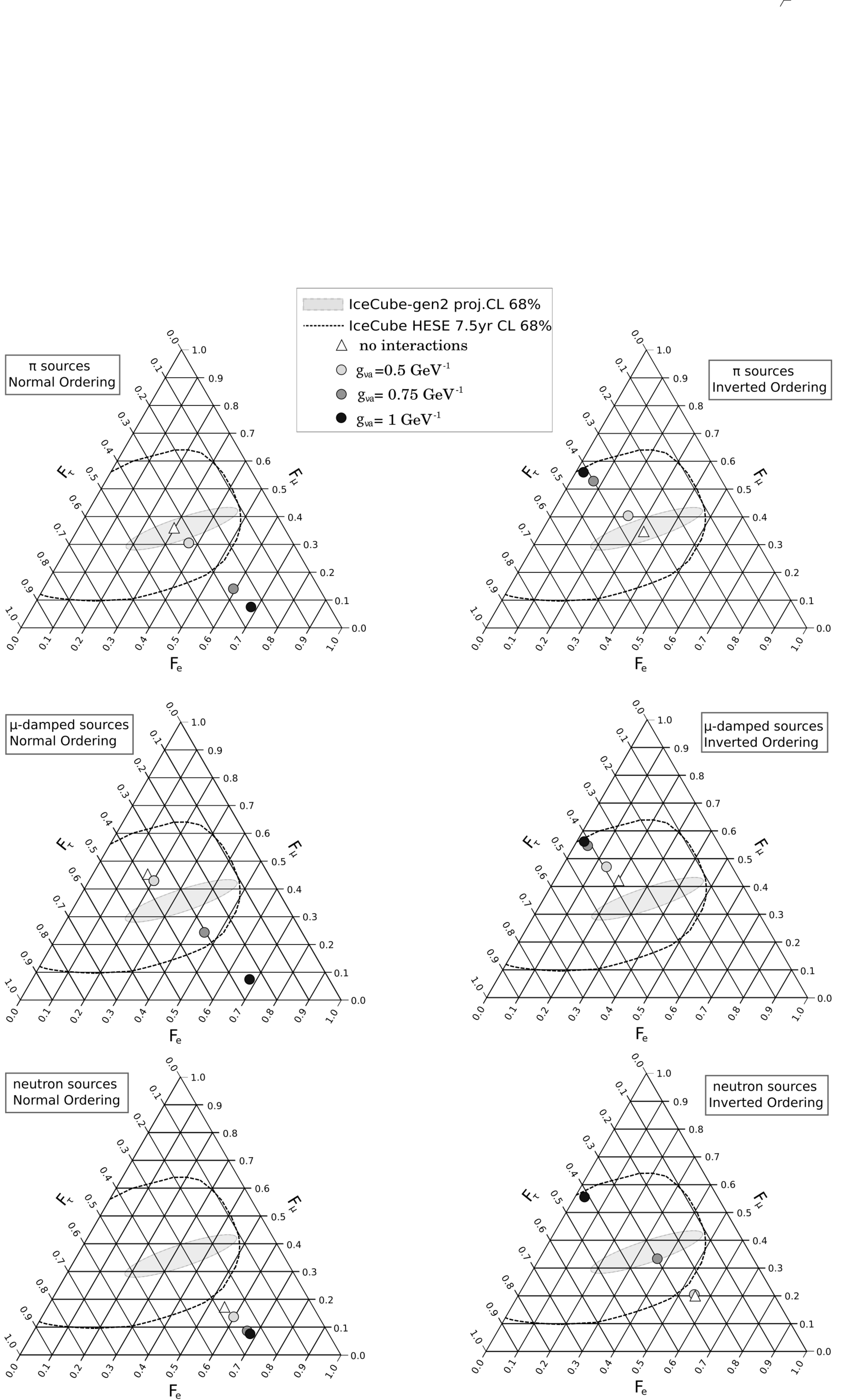}
\caption{Ternary plots of the flavor ratios for neutrinos above $E_{\rm min}=5\times 10^5{\rm GeV}$ in the NO case (left plots) and in the IO case (right plots) under the assumptions of pion decay sources (upper panels), muon damped sources (middle panels), and neutron decay sources (bottom panels). The outcomes expected in the absence of neutrino interactions are marked by white triangles, while the results with interactions enabled are shown in gray squares for $g_{\nu a}=0.5\,{\rm GeV}^{-1}$, in dark gray squares for $g_{\nu a}=0.75\,{\rm GeV}^{-1}$, and in black for $g_{\nu a}=1\,{\rm GeV}^{-1}$. For reference, we show with black-dashed lines the contour corresponding to the $68\%$ C.L. obtained by IceCube with 7.5yr of accumulated data of the HESE sample \cite{stettner2019}, and we show in gray the projected region expected to be obtained with IceCube-gen2 if the sources were dominated by pion decay.}\label{fig11-ternary-plot}   
\end{center}
\end{figure*}

In fact, this can be illustrated with a ternary plot as the one in Fig. \ref{fig11-ternary-plot}, where each position represents a unique combination of the three flavor ratios obtained by integration of the  neutrino fluxes over the energy and also over all incoming directions. In this plot, we mark {with a white triangle} the expected values in the cases of no interactions corresponding to pion decay (top panels), muon damped (middle panels), and neutron decay sources (bottom panels). If the coupling is set as $g_{\nu a}=0.5\,{\rm GeV}^{-1}$ then the results shift to the indicated by the gray squares, to the dark gray ones if we set $g_{\nu a}=0.75\,{\rm GeV}^{-1}$, and to the black ones for $g_{\nu a}=1\,{\rm GeV}^{-1}$. We also include the $68\%$ CL region obtained with the sample of high energy starting events (HESE) by IceCube \cite{stettner2019}, and the projected $68\%$ CL constrained ragion with 8 yr of IceCube if the composition were that for pion decay sources, i.e., $(f_{e,\rm s}:f_{\mu,\rm s}:f_{\tau,\rm s})=(2:1:0)$ (see Ref. \cite{icecubegen2}).

\section{Discussion}\label{sec5:discussion}
In this work, we have analyzed the impact of possible interactions of astrophysical neutrinos with UALP DM particles $a$, with $m_a\approx 5\times 10^{-22}{\rm eV}$. Taking into account a diagonal coupling with the same value for the three flavors ($g_{\nu_\alpha a}=g_{\nu a}$), we treated the extragalactic and galactic propagation using transport equations for the density of each massive neutrino. 
Considering as benchmarks three usually adopted initial flavor compositions for pion decay, muon-damped , and neutron decay sources, we found that interactions can cause an important change in the flavor composition to be observed on Earth with neutrino telescopes, and this effect is also sensible to the the mass ordering of the massive neutrinos. As the coupling is increased gradually from $g_{\nu a}=0.5\,{\rm GeV}^{-1}$ to $g_{\nu a}=1\,{\rm GeV}^{-1}$, the flavor composition gradually departs from the result corresponding to no interactions to $(F_e:F_\mu:F_\tau)_{\rm NO}= (|U_{e1}|^2:|U_{\mu 1}|^2:|U_{\tau 1}|^2)$ in the NO case, and to $(F_e:F_\mu:F_\tau)_{\rm IO}= (|U_{e3}|^2:|U_{\mu 3}|^2:|U_{\tau 3}|^2)$ in the IO case. We also found that for somewhat smaller couplings $0.5\, {\rm GeV}^{-1}\lesssim g_{\nu a}\lesssim 1\, {\rm GeV}^{-1}$, neutrinos are affected differently at different directions in the sky because the column of galactic DM will be higher for neutrino path passing closer to the galactic center. Therefore, if the measured neutrino flavor composition is homogeneously distributed across the whole sky, this will be evidence against neutrino interactions with UALP DM in the mentioned range of the coupling $g_{\nu a}$. 
Considering the current data from IceCube regarding the flavor composition \cite{stettner2019}, the neutrino interactions discussed in this work with a coupling $g_{\nu a}\gtrsim 1\, {\rm GeV}^{-1}$ appear to be disfavored, as can be seen in the ternary plots of Fig. \ref{fig11-ternary-plot}. Still, more statistics is necessary to obtain a higher precision in the experimental determination of the flavor ratios and this will be achieved with longer time exposure and larger detectors such as IceCube-gen2 \cite{icecubegen2}.  

These conclusions are in agreement with the constraint estimated in Ref.\cite{huang2018} based on the detection of the neutrino event IceCube-170922A, leading to $g_{\nu a}\gtrsim 0.8{\rm GeV}^{-1}$, and in particular our analysis  shows that a lower coupling $g_{\nu a}=0.5{\rm GeV}^{-1}$ interactions leave their signature by causing a change in the flavor ratios for energies greater than $\sim 10^5{\rm GeV}$, with respect to the result corresponding to no interactions, as shown in the left panels of Figs. \ref{fig7-flavor-ratios-E-NO} and \ref{fig8-flavor-ratios-E-IO}. 

 Given that the energy dependence of the flavor ratios is expected to be experimentally determined by future measurements \cite{icecubegen2,bustamante2021}, the analysis here presented can be useful to probe a {diagonal coupling with a value $g_{\nu a}\approx 0.5{\rm GeV}^{-1}$ for all the flavors. This is below the ranges that, as mentioned in Ref. \citep{huang2018}, can be excluded by avoiding a too fast cooling in supernovae ($g_{\nu a}\lesssim 10^4{\rm GeV^{-1}}$) \citep{farzan2003} and also by allowing the free streaming of neutrinos in the early universe before photon decoupling ($g_{\nu a}\lesssim 10^3{\rm GeV^{-1}}$)\citep{hannestad2005}\footnote{We note that the mentioned bound applies, since it refers to the case of diagonal couplings in the mass base, which is also our case according to Eq. (\ref{lagr2}). }. 
 In the cases of different couplings for each flavor or even non-diagonal couplings, observations with upcoming underground neutrino detectors such as JUNO \cite{juno2015} and DUNE \cite{dune2015} are expected to explore values as $g_{\nu_\alpha a}\sim 0.1-0.01\,{\rm GeV}^{-1}$ searching for oscilation effects \cite{huang2018}.   }

We leave for a future work the treatment of higher energy neutrino fluxes such as the so-called comogenic neutrinos produced by cosmic ray interactions with the cosmic microwave background. Given that the energy loss per interaction becomes higher at higher energies, the interactions of these neutrinos is therefore expected to alter the flavor composition for lower $g_{\nu a}$ values than the explored in the present work.

\begin{acknowledgements}
  We thank CONICET and Universidad Nacional de Mar del Plata for their financial support through grants PIP 0046 and  15/E870EXA912/18, respectively.
\end{acknowledgements}

\end{document}